\documentclass[twocolumn,showpacs,preprintnumbers,amsmath,amssymb,prb]{revtex4}
\usepackage{graphicx}
\usepackage[colorlinks=true,linkcolor=blue,citecolor=blue]{hyperref}
\usepackage{color}

\begin{document}


\title{Low-energy effective theory and two distinct critical phases in a spin-$\frac{1}{2}$ frustrated three-leg spin tube}

\author{Yang Zhao$^{1}$, Shou-Shu Gong$^{2}$, Yong-Jun Wang$^{3}$, Gang Su$^{1}$}
\email[Author to whom correspondence should be addressed. ]{Email:
gsu@ucas.ac.cn}
 \affiliation{$^{1}$Theoretical Condensed Matter Physics and Computational
Materials Physics Laboratory, School of Physics, University of Chinese Academy
of Sciences, P. O. Box 4588, Beijing 100049, China
\linebreak $^{2}$Department of Physics and Astronomy, California State University, Northridge, California 91330, USA
\linebreak $^{3}$Department of Physics, Beijing Normal University, Beijing 100875, China
}

\date{\today}

\begin{abstract}
Motivated by the crystal structures of [(CuCl$_{2}$tachH)$_{3}$Cl]Cl$_{2}$ and Ca$_{3}$Co$_{2}$O$_{6}$, we develop a low-energy effective theory using the bosonization technique for a spin-1/2 frustrated three-leg spin tube with trigonal prism units in two limit cases. The features obtained with the effective theory are numerically elucidated by the density matrix renormalization group method. Three different quantum phases in the ground state of the system, say, one gapped dimerized phase and two distinct gapless phases, are identified, where the two gapless phases are found to have the conformal central charge $c = 1$ and $3/2$, respectively. Spin gaps, spin and dimer correlation functions, and the entanglement entropy are obtained. In particular, it is disclosed that the critical phase with $c =3/2$ is the consequence of spin frustrations, which might belong to SU(2)$_{k=2}$ Wess-Zumino-Witten-Novikov universality class, and is induced by the twist term in the bosonized Hamiltonian density.

\end{abstract}

\maketitle

\section{\label{sec:level1}Introduction}
\begin{figure}[tbp]
  \centering
  \includegraphics[width=0.8\linewidth,clip]{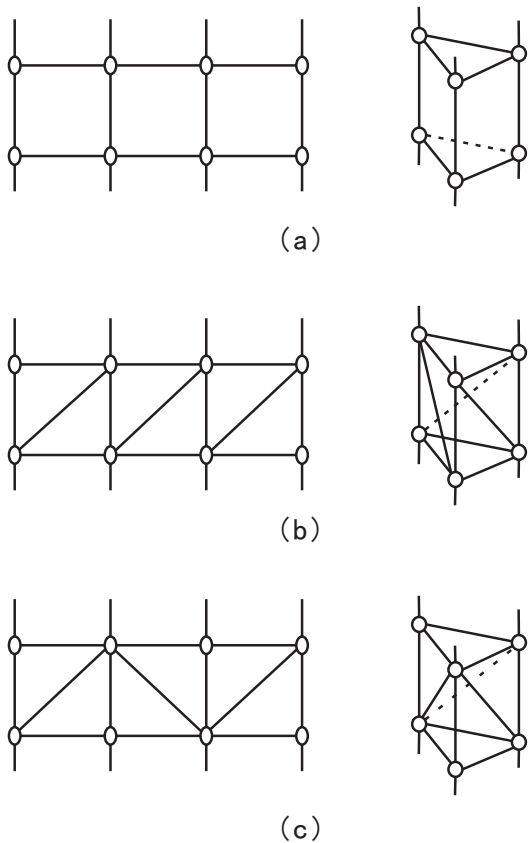}\\
  \caption{Spin ladders and its corresponding spin tubes (the ladder with periodical condition along the rung direction). (a) shows the standard spin ladder and the typical spin tube. (b) and (c) are the ladders with next nearest couplings and the spin tubes generated by them. }\label{prestruc}
\end{figure}

With the discovery of several spin tube materials, such as [(CuCl$_{2}$tachH)$_{3}$Cl]Cl$_{2}$ \cite{tubecu}, Na$_{2}$V$_{3}$O$_{7}$ \cite{tubena} and Cu$_{2}$Cl$_{4}\cdot$ D$_{8}$C$_{4}$SO$_{2}$ \cite{tubed}, etc., much effort
has been made to investigate the low-lying excitations \cite{Tube gapa, Tube gapb, Dimerization, frustubea, low lying e, deconfined}, long range order \cite{vec chir a, vec chir b}, phase transitions \cite{Tube gapb, frustubeb}, magnetization plateau \cite{plata, mag tube, Zhao} and other ground state properties of spin tubes due to the inter-chain frustrated couplings. A spin tube, geometrically, can be seen as a multi-leg spin ladder with periodic condition along the rung direction (See Fig. \ref{prestruc} for the case of three-leg ladders). Therefore, the antiferromagnetic (AF) inter-chain couplings in the odd-leg spin tubes usually give rise to geometrical frustrations. For the typical spin tubes \cite{Tube gapa, Tube gapb, Dimerization} [Fig. \ref{prestruc} (a)], the excitation is gapped when the inter-chain couplings are identical, and becomes gapless when the interactions are different to some certain extent from each other. The opening and closing of the gap correspond to a dimerized phase and a critical phase with central charge $c= 1$, respectively.
Recently, a few new spin tubes with more complex inter-chain couplings \cite{frustubea, frustubeb, Zhao, CaCo} [Fig. \ref{prestruc} (b) and (c))] have attracted much attention, in which frustrated inter-chain couplings could generate twist terms that may lead to unknown quantum phases in the ground state of the system. The twist operator is first found in the effective low energy model of a zigzag spin ladder \cite{RGE} owing to the frustrated inter-chain couplings. This parity-breaking operator is marginal in the renormalization group (RG) sense and has a conformal spin 1\cite{Tsvbook}. Currently, the twist operator is believed to be the origin of incommensurate correlations in XXZ zigzag spin ladders \cite{RGE, Tesv twis, zig coma, zig comb, zigzag}. For an isotropic zigzag spin ladder, the twist operator is proposed to cause dimerization \cite{RGE} in the ferromagnetic inter-chain coupling region. However what role this operator plays in a spin tube is still unclear, which needs a further study.

\begin{figure}[tbp]
\includegraphics[width=0.8\linewidth,clip]{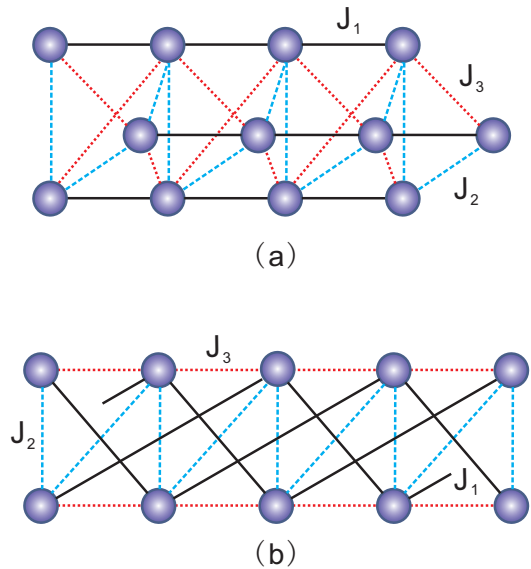}
\caption{(Color online) (a) The structure of the frustrated three-leg spin tube under investigation. The circles (purple) indicate the sites with spin $S$ = $\frac{1}{2}$. The antiferromagnetic couplings $ \emph{J}_{2} $ and $ \emph{J}_{3} $ are presented by
dashed line (blue) and dotted line (red), respectively. The
solid line (black) indicates the AF interaction along the leg direction. (b) shows the equivalent spin ladder structure transformed from (a) by setting $ \emph{J}_{3} $ bonds as legs.} \label{Structure}
\end{figure}

In this work, by means of the bosonization technique and the density matrix renormalization group (DMRG) method we shall study a frustrated three-leg spin tube comprised of trigonal prism units, which derives from the crystal structures of [(CuCl$_{2}$tachH)$_{3}$Cl]Cl$_{2}$ \cite{tubecu} and Ca$_{3}$Co$_{2}$O$_{6}$ \cite{CaCo}, as shown in Fig. \ref{Structure} (a), where the AF couplings $\emph{J}_{2}$ and $\emph{J}_{3}$ form helical paths along the tube direction. Such a spin tube can be transformed to an equivalent spin ladder, as shown in Fig. \ref{Structure} (b). Owing to the complex zigzag-like inter-chain couplings, the twist operator in the bosonized Hamiltonian should depend on both $J_{2}$ and $J_{3}$, where we take the coupling $J_{1}$ as an energy scale. Consequently, we only need to adjust $\emph{J}_{2}$ and $\emph{J}_{3}$ to study the effect of the twist term on the properties in the ground state. For the sake of simplicity, we consider the case of spin $S=1/2$. Interestingly, we showed that a new critical phase with central charge $c = \frac{3}{2}$ appears in this system, which is found in spin tubes for the first time and reveals a novel physical effect of the twist operator. In addition, one dimer phase and one conventional critical phase with central charge $c = 1$ due to the competition of the AF couplings $\emph{J}_{2}$ and $\emph{J}_{3}$ are also identified in the ground state.

\section{Bosonization and low-energy effective theory}
The Hamiltonian of the present spin tube could be written as
\begin{eqnarray}
H &=&\sum_{i=1}^{L/3}[\emph{J}_{1}(\mathbf{S}_{3i-2}\cdot \mathbf{S}_{3i+1}+%
\mathbf{S}_{3i-1}\cdot \mathbf{S}_{3i+2}+\mathbf{S}_{3i}\cdot \mathbf{S}%
_{3i+3})
\nonumber \\
&+&\emph{J}_{2}(\mathbf{S}_{3i-1}\cdot \mathbf{S}_{3i+1}+\mathbf{S}%
_{3i}\cdot \mathbf{S}_{3i+1}+\mathbf{S}_{3i}\cdot \mathbf{S}_{3i+2})
\nonumber \\
&+&\emph{J}_{3}(\mathbf{S}_{3i-2}\cdot \mathbf{S}_{3i-1}+\mathbf{S}%
_{3i-1}\cdot \mathbf{S}_{3i}+\mathbf{S}_{3i-3}\cdot
\mathbf{S}_{3i+1})], \label{ham}
\end{eqnarray}
where $\mathbf{S}_{j}$ is the spin operator on the \emph{j}th site, $L$ is
the total number of sites, and $\emph{J}_{1}$,
$\emph{J}_{2}$, $\emph{J}_{3}>0$ are all AF couplings. In what follows we consider two limit cases: (I) $\emph{J}_{2} \gg \emph{J}_{3},\emph{J}_{1}$, and (II) $\emph{J}_{3} \gg \emph{J}_{2}, \emph{J}_{1}$. In the following DMRG calculations $\emph{J}_{1}$ is set to be unity for simplicity.

\subsection{$\emph{J}_{2} \gg \emph{J}_{3}, \emph{J}_{1}$}

This case is quite simple. From Fig. \ref{Structure}, one may note that the system can be regarded as a single spin chain with next nearest and next next nearest neighbor interactions. The Hamiltonian can be written as
 \begin{eqnarray}
 H &=& \emph{J}_{2}\sum_{i}\mathbf{S}_{i}\cdot\mathbf{S}_{i+1} + \emph{J}_{3}\sum_{i}\mathbf{S}_{i}\cdot\mathbf{S}_{i+2} \nonumber\\
 &+& \emph{J}_{1}\sum_{i}\mathbf{S}_{i}\cdot\mathbf{S}_{i+3}, \label{eqc1} 
 \end{eqnarray}
 where the last two terms can be viewed as perturbations. Following the standard bosonization techniques \cite{Affleck}, in the continuum limit, the Hamiltonian density of Eq. (\ref{eqc1}) can be written as:
\begin{equation}
    \mathcal{H}=\mathcal{H}_{0}+\mathcal{H}_{JJ}, \label{bosonc1}
 \end{equation}
\begin{eqnarray}
 \mathcal{H}_{0}=\frac{2\pi v_{s}}{3}(:\mathbf{J}_{L}\cdot \mathbf{J}_{L}:+:\mathbf{J}_{R}\cdot \mathbf{J}_{R}:), 
 \end{eqnarray}
 \begin{equation}
    \mathcal{H}_{JJ}=g\mathbf{J}_{L}\cdot \mathbf{J}_{R},
 \end{equation}
 \begin{equation}
    g=2\emph{J}_{3}+4\emph{J}_{1} - \emph{J}_{c},
 \end{equation}
 where $\mathbf{J}_{L,R}$ are left or right moving SU(2) current operators representing the smooth magnetization part of spin density operators \cite{GNT}, $v_{s}\sim\emph{J}_{2}a_{0}$ is the spin velocity with $a_{0}$ the lattice constant, and $\emph{J}_{c}>$0. Obviously, Eq. (\ref{bosonc1}) has the same form as the Hamiltonian density for a zigzag spin ladder \cite{zigzag} except for the coefficient $g$ of $\mathcal{H}_{JJ}$ term. When $g$$>$0, the system is in the dimer phase with an energy gap:
  \begin{equation}
    \Delta \simeq \textrm{exp}(-\frac{const}{g}).
 \end{equation}
 It is deserved to mention here that this gap could also survive even for $J_1$$\gg$$J_2$,$J_3$, in which the system can be seen as the three free spin chains with intra-chain couplings $J_1$ perturbed by relevant perturbations generated by the inter-chain couplings $J_2$ and $J_3$, as in the case of spin ladders\cite{GNT, Cabra, ALLen}. Therefore, this case as well falls into the gapped phase.
 At last, when $g$$<$0 it is in the Luttinger liquid phase with central charge $c$=1. The phase transition between the two phases is of a Kosterlitz-Thouless transition \cite{Tube gapb, GNT}.

\begin{figure}[tbp]
\includegraphics[width = 0.8\linewidth,clip]{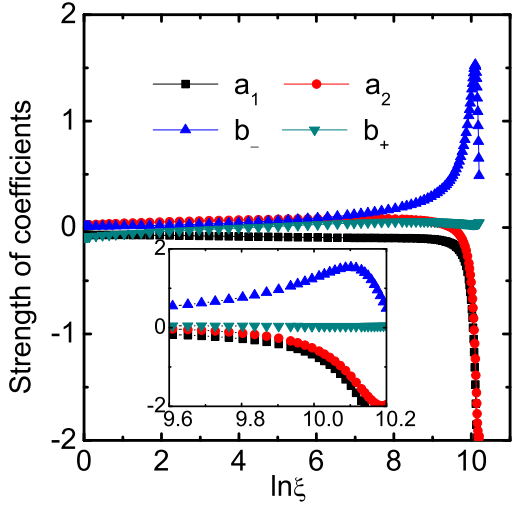}
\caption{(Color online) The renormalization group (RG) flows for the three-leg frustrated spin tube under interest with $\emph{J}_{3}$ $\gg$ $\emph{J}_{2}$. The variable $b_{\pm}$ = $\frac{b_{2}\pm b_{1}}{2}$ is the same as the definition of Ref. \onlinecite{RGE}. The inset is the enlarged part where the RG flows to the strong coupling limit can be seen clearly. $\xi$ is the logarithmic variable in the renormalization group equations.}
\label{RGFLOW}
\end{figure}

\subsection{$\emph{J}_{3} \gg \emph{J}_{2}, \emph{J}_{1}$}

This case is very interesting. In this situation, the model can be reshaped into a spin ladder structure as shown in Fig. \ref{Structure} (b). By treating all inter-chain couplings as perturbations, after the bosonization procedure, the corresponding Hamiltonian density can be written as:
 \begin{equation}
    \mathcal{H}=\mathcal{H}_{0}+\mathcal{H}_{LR}+\mathcal{H}_{JJ}+\mathcal{H}_{twist}, \label{bosonc2}
 \end{equation}
\begin{eqnarray}
  \mathcal{H}_{0} &=& \frac{2\pi v_{s}}{3}\sum_{j=1,2}(:\mathbf{J}_{j,L}\cdot\mathbf{J}_{j,L}:+:\mathbf{J}_{j,R}\cdot\mathbf{J}_{j,R}:),
\end{eqnarray}
\begin{equation}
    \mathcal{H}_{LR}=g_{LR}(\mathbf{J}_{1,L}\cdot\mathbf{J}_{1,R}+\mathbf{J}_{2,L}\cdot\mathbf{J}_{2,R}),
\end{equation}
\begin{equation}
    \mathcal{H}_{JJ}=2(\emph{J}_{1}+\emph{J}_{2})(\mathbf{J}_{1,L}+\mathbf{J}_{1,R})\cdot(\mathbf{J}_{2,L}+\mathbf{J}_{2,R}),
\end{equation}
\begin{equation}
    \mathcal{H}_{twist}=-\frac{3\emph{J}_{1}-\emph{J}_{2}}{2}(\mathbf{n}_{1}\partial_{x}\mathbf{n}_{2}-\mathbf{n}_{2}\partial_{x}\mathbf{n}_{1}),\label{twi}
\end{equation}
where $\mathbf{n}_{i}$ is the staggered part of the spin density operator with the subscript $i$ the leg index, spin velocity $v_s$$\sim$$J_3a_0$, and $\mathbf{J}_{j,L,R}$ is the left or right current operator of the $j$th leg. The coefficient $g_{LR}$ $\propto$ $-\emph{J}_{3}$, and $\mathcal{H}_{LR}$, $\mathcal{H}_{JJ}$ and $\mathcal{H}_{twist}$ are marginal perturbations with scaling dimension $d$ = 1 \cite{RGE}. $\mathcal{H}_{twist}$ is the twist term that is produced by the frustrated inter-chain interactions. As pointed out in Refs. [\onlinecite{RGE, Tesv twis}], it arouses a spin nematic phase in the ground state of the XXZ zigzag spin ladder. Its effect on an isotropic Heisenberg model is still not very clear and needs further investigations.

In fact, this present case provides a convenient way to study the effect of the twist term on an isotropic Heisenberg system since the coefficient of $\mathcal{H}_{twist}$ could be adjusted by regulating the value of $\emph{J}_{2}$. In order to look at the effect of a group of marginal operators, we resort to solve the renormalization group equations \cite{RGE, Cabra} numerically.

First, following the procedure of Refs. [\onlinecite{RGE, GNT, ALLen}], we can rewrite $\mathcal{H}_{LR}$, $\mathcal{H}_{JJ}$ and $\mathcal{H}_{twist}$ using Majorana fermions $\zeta^{0,1,2,3}_{L,R}$ as
\begin{eqnarray}
  \mathcal{H}_{pert} &=& a_{1}(\zeta^{0}_{R}\zeta^{1}_{L}\zeta^{2}_{L}\zeta^{3}_{L}+\zeta^{0}_{L}\zeta^{1}_{R}\zeta^{2}_{R}\zeta^{3}_{R})\nonumber\\
  &+& a_{2}(\zeta^{0}_{R}\zeta^{1}_{R}\zeta^{2}_{R}\zeta^{3}_{L}+\zeta^{0}_{L}\zeta^{0}_{L}\zeta^{0}_{L}\zeta^{0}_{R})\nonumber \\
  &+& a_{3}(\zeta^{0}_{R}\zeta^{1}_{R}\zeta^{2}_{L}\zeta^{3}_{R}+\zeta^{0}_{R}\zeta^{1}_{L}\zeta^{2}_{R}\zeta^{3}_{R}\nonumber\\
  &+& \zeta^{0}_{L}\zeta^{1}_{L}\zeta^{2}_{R}\zeta^{3}_{L}+\zeta^{0}_{L}\zeta^{1}_{R}\zeta^{2}_{L}\zeta^{3}_{L}) \nonumber \\
  &+& b_{1}(\zeta^{0}_{R}\zeta^{0}_{L}\zeta^{1}_{R}\zeta^{1}_{L}+\zeta^{0}_{R}\zeta^{0}_{L}\zeta^{2}_{R}\zeta^{2}_{L}+\zeta^{0}_{R}\zeta^{0}_{L}\zeta^{3}_{R}\zeta^{3}_{L})\nonumber\\
  &+& b_{2}(\zeta^{1}_{R}\zeta^{1}_{L}\zeta^{2}_{R}\zeta^{2}_{L}+\zeta^{1}_{R}\zeta^{1}_{L}\zeta^{3}_{R}\zeta^{3}_{L}+\zeta^{2}_{R}\zeta^{2}_{L}\zeta^{3}_{R}\zeta^{3}_{L}),\nonumber\\
\end{eqnarray}
where $a_{1} = -\frac{1}{3}a_{2} = \frac{1}{3}a_{3} = \gamma (-\frac{3\emph{J}_{1}-\emph{J}_{2}}{2})$, $\gamma$ is a nonzero constant and $b_{1} = -\frac{g_{LR}}{2}-\emph{J}_{1}-\emph{J}_{2}$, $b_{2} = -\frac{g_{LR}}{2}+\emph{J}_{1}+\emph{J}_{2}$. The terms with the coefficient $a_{i}$ stem from $\mathcal{H}_{twist}$ and $b_{i}$ from both $\mathcal{H}_{LR}$ and $\mathcal{H}_{JJ}$.

Next, we set $\emph{J}_{2}$ $<$ $3\emph{J}_{1}$ to keep the coefficient of $\mathcal{H}_{twist}$ negative that cannot be achieved in a usual zigzag Heisenberg spin ladder. Afterwards, choosing the initial values of $a_{i}$ and $b_{i}$ in the range where $\emph{J}_{3}$ $\gg$ $\emph{J}_{2}, \emph{J}_{1}$, we obtain a new solution of the renormalization group equations \cite{RGE} as shown in Fig. \ref{RGFLOW}, where the RG flows corresponding to $\mathcal{H}_{twist}$ go to the strong coupling regime first, which suggests a new phase may come into being. Recall that for a two-leg zigzag spin ladder, the RG flows corresponding to $\mathcal{H}_{JJ}$ go to the strong coupling regime first. In the following we shall examine numerically the results from the low-energy effective theory.

\section{Spin gap, correlations and entanglement entropy}

To substantiate the above bosonization analysis, we use the DMRG method\cite{DMRG} to numerically calculate the spin gap, spin-spin and dimer-dimer correlation functions, as well as entanglement entropy of the model. In the calculations, up to $8000$ optimal states are kept, and the truncation errors are of $10^{-12}$ for the ground state and $10^{-7}$ for the excited states.

\begin{figure}
\centering
\includegraphics[width=1.5\linewidth,clip]{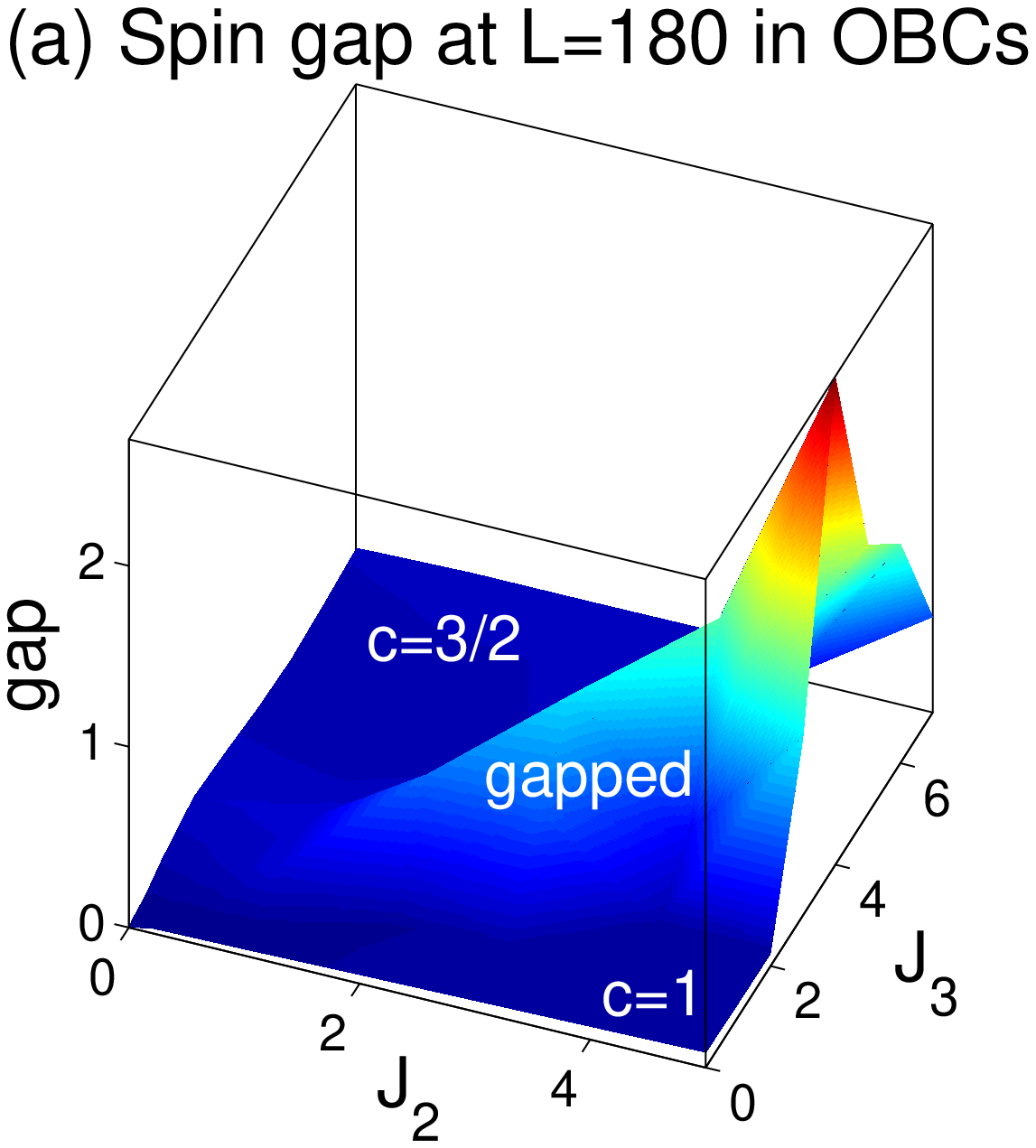}
\includegraphics[width=0.6\linewidth,clip]{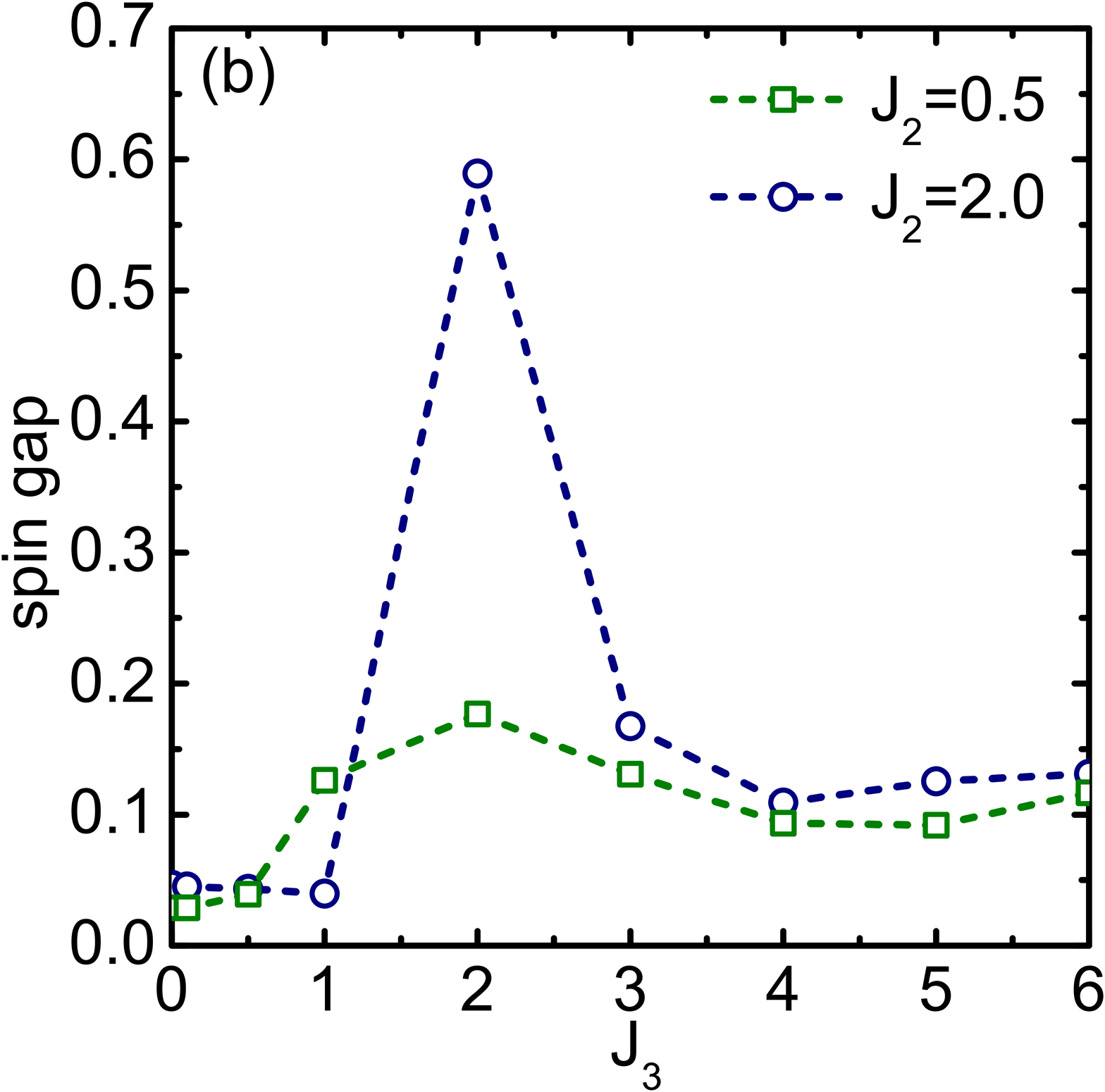}
\includegraphics[width=1.0\linewidth,clip]{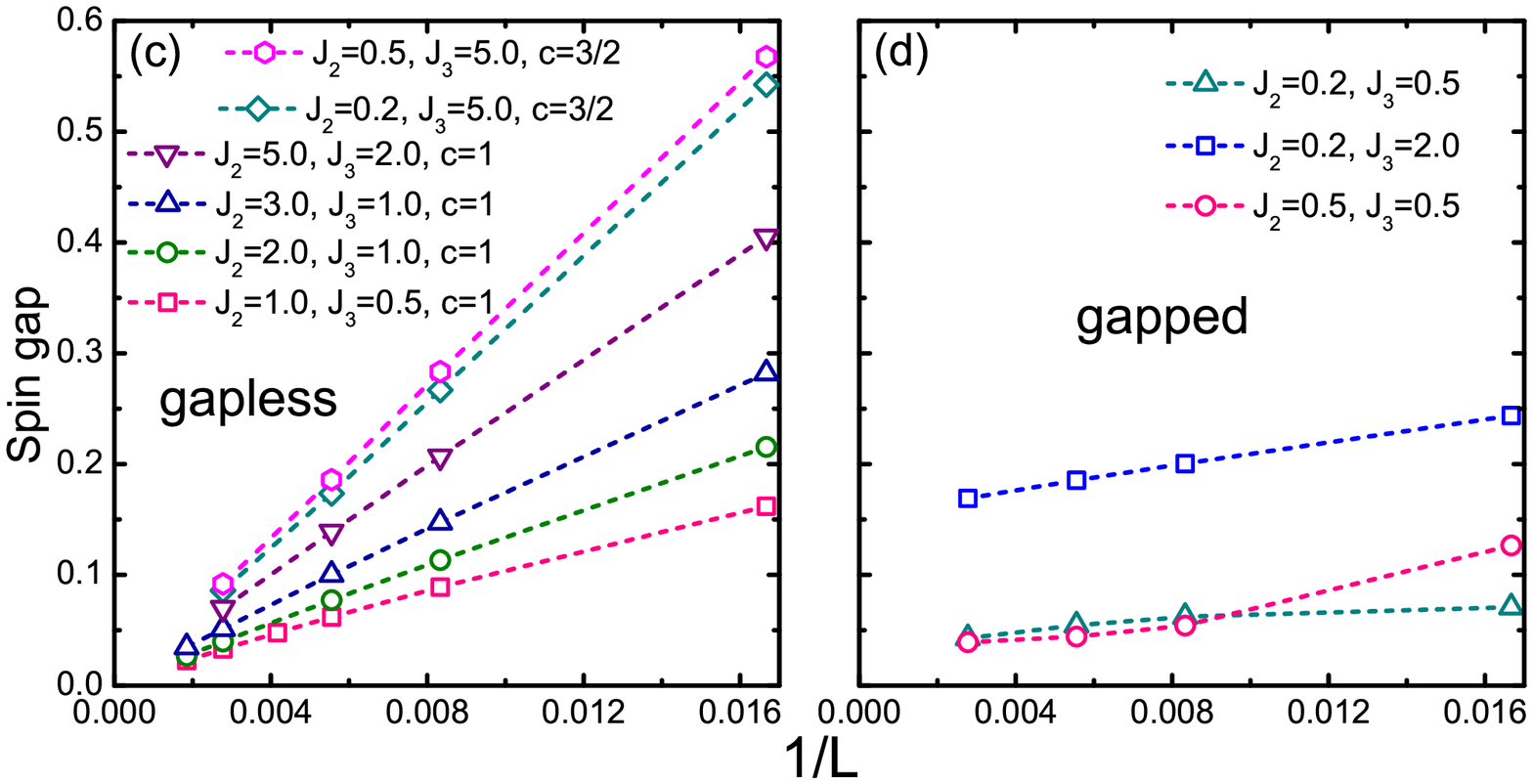}
\caption{(Color online) (a) Coupling dependence of the spin gap at $L$=$180$ obtained with OBCs. (b) $J_3$ dependence of the spin gap for $J_2=0.5$ and $2.0$. Finite-size scaling of spin gap for (c) the gapless and (d) the gapped phases.} \label{gapproj}
\end{figure}

\subsection{Spin gap}

Figure  \ref{gapproj}(a) is the coupling dependence of the finite-size spin gap at $L$=$180$ obtained in open boundary conditions (OBCs). With increasing $J_3$ for any $J_2$, the spin gap has two area with tiny values relatively, between which there is a raised region of spin gap. In Fig. \ref{gapproj}(b), $J_3$ dependence of spin gap at $J_2$=$2.0$ is shown, where the spin gaps increase to the maximum value in the raised region and decrease slowly with growing $J_3$, while those in other regions diminish fast, probably suggesting a gapped spin excitation in the middle region and a gapless excitation in other regimes. As shown in Figs. \ref{gapproj}(c) and (d), the spin gaps away from the raised region extrapolate to zero, while those within the raised region go to finite values in the thermodynamic limit, therefore indicating two gapless and one gapped phases. These results are also in accordance with our previous bosonization analysis that the system will be in the gapless phase for $J_2$$\gg$$J_3$, $J_1$ and $J_2$, $J_1$$\ll$$J_3$, and is gapped in other parameter regimes.

\subsection{Spin-spin and dimer-dimer correlation functions}

\begin{figure}
\includegraphics[width=1.0\linewidth,clip]{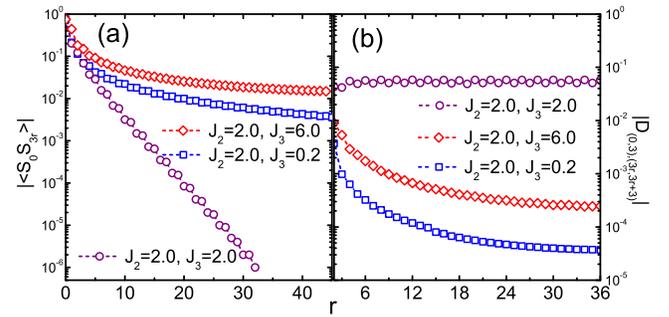}
\caption{(Color online) (a) The spatial dependence of the spin-spin correlation function $\langle \mathrm{S}_{0}\cdot\mathrm{S}_{3r}\rangle$ in the three phases. For $J_3$=$0.2$ and $2.0$, the spins ($\mathrm{S}_{3r}$) are assumed along the leg direction as indicated in Fig. \ref{Structure}(a), while for $J_3$=$6.0$, the spins along the leg as indicated in Fig. \ref{Structure}(b). (b) The spatial dependence of the dimer correlation $|D_{(0,3),(3r,3r+3)}|$ in the three phases. The dimers are supposed along the leg shown in Fig. \ref{Structure}(a).} \label{dimerph}
\end{figure}

The spin-$\frac{1}{2}$ 1D gapless spin systems usually preserve the translation symmetry and have the spin correlations with power-law decay, while the gapped ones should break the translation symmetry and have spin correlations of exponential decay. Figure \ref{dimerph} (a) presents the spin correlation function $\vert\langle\mathbf{S}_{0}\cdot\mathbf{S}_{3r}\rangle\vert$ in different phases. For $J_2$=$2.0$, $J_3$=$0.2$ and $J_2$=$2.0$, $J_3$=$6.0$, which are in the two gapless phases, $\vert\langle\mathbf{S}_{0}\cdot\mathbf{S}_{3r}\rangle\vert$ decays with a power law, while at $J_2$=$J_3$=$2.0$ that is in the gapped phase, $\vert\langle\mathbf{S}_{0}\cdot\mathbf{S}_{3r}\rangle\vert$ has an exponential decay with a short correlation length $\xi \simeq 2.7$. The breaking of translational symmetry is usually detected by the dimer-dimer correlation
function $D_{(i,j),(k,l)}$=$\langle(\mathbf{S}_{i}\cdot\mathbf{S}_{j})(\mathbf{S}_{k}\cdot\mathbf{S}_{l})\rangle-\langle \mathbf{S}_{i}\cdot\mathbf{S}_{j}\rangle \langle \mathbf{S}_{k}\cdot\mathbf{S}_{l}\rangle$. In Fig. \ref{dimerph}(b), $|D_{(0,3),(3r,3r+3)}|$ is shown for the three phases. In both gapless phases, the dimer correlations decay with a power law, which is consistent with the preserved translation symmetry and gapless spin excitations. In the gapped phase, the dimer correlation builds a long-range order (LRO).

\begin{figure}

\includegraphics[width=0.8\linewidth,clip]{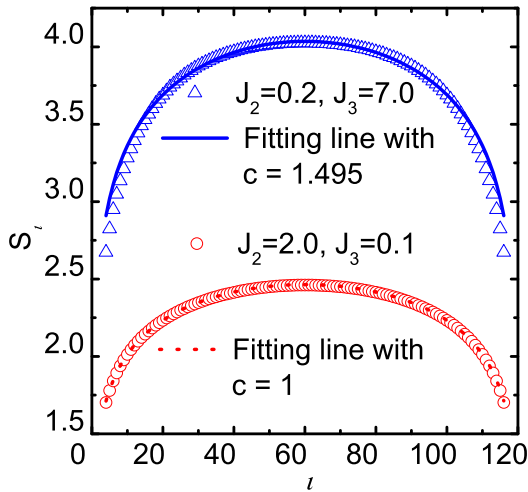}
\caption{(Color online) Fitting of the entanglement entropy given by Eq. (\ref{entropy_c}) to the DMRG results in the two gapless phases. The central charges from the fitting are obtained to be $c$=$1$ and $3/2$, respectively.}\label{fentropy}
\end{figure}

\subsection{Entanglement entropy}

To distinguish the two gapless phases\cite{spinbose}, we calculated the central charge $c$ from the block entanglement entropy of the system $S_{l}$=$-\mathrm{Tr}\rho_{l}\ln\rho_{l}$, where $\rho_{l}$ is the reduced density matrix of a subsystem of size $l$. For a gapless system with periodic boundary conditions (PBCs), the entanglement entropy is given by
\begin{equation}
S_{l}=\frac{c}{3}\ln\left[\frac{N}{\pi}\sin\left(\frac{\pi l}{N}\right)\right]+g_{\mathrm{PBCs}}, \label{entropy_c}
\end{equation}
where $N$ is the total length of the system, and $g_{\mathrm{PBCs}}$ is a nonuniversal constant with PBCs \cite{entropy}. The central charge is obtained by fitting the entanglement entropy given by Eq. (\ref{entropy_c}) to the DMRG results for the system with $L$ up to $360$.

As shown in Fig. \ref{fentropy}, in the gapless phase with large $J_2$ (for instance $J_2$=$2.0$, $J_3$=$0.1$), both results for $S_{l}$ are fitted quite well with $c$=$1$, which indicates that this gapless phase belongs to the same universality class as the spin-$\frac{1}{2}$ Heisenberg AF chain. In the gapless phase with large $J_3$ like $J_2$=$0.2$, $J_3$=$7.0$, the central charge is identified as $c$=$\frac{3}{2}$. Therefore, we can label the different phases with its central charge in Fig. \ref{gapproj} (a), which comprises the phase diagram.  As the model preserves the SU(2) symmetry, the transition from the dimer phase to the gapless phase with $c$=$\frac{3}{2}$ might be in the SU(2)$_{k=2}$ Wess-Zumino-Witten-Novikov (WZWN) universality class \cite{Affleck}. In combination with the bosonization analysis, it is observed that for $J_3$$\gg$$J_2$ the RG flows of $\mathcal{H}_{twist}$ go to the strong coupling regime faster than $\mathcal{H}_{JJ}$ that leads to a dimerized phase \cite{RGE,GNT}, and one may judge that such a gapless phase with nontrivial central charge $c$=$\frac{3}{2}$ in the frustrated three-leg Heisenberg spin tube is probably induced by the twist term $\mathcal{H}_{twist}$.

\section{Summary}

To summarize, by means of the bosonization technique we develop a low-energy effective theory for the spin-1/2 frustrated three-leg spin tube in two limit cases, and also invoke the DMRG calculations on the spin gap, spin-spin and dimer-dimer correlation functions as well as the entanglement entropy to elucidate the effective analyses. We have discovered a dimer phase and two distinct critical phases in this system. The dimer phase is characterized by the finite spin gap, exponentially decaying spin-spin correlations, and a dimer-dimer LRO. The critical phases are found to have gapless spin excitations, power-law decaying spin and dimer correlations. The different central charges $1$ and $\frac{3}{2}$ distinguish the two critical phases. Based on the bosonization analysis, the novel critical phase with $c$=$\frac{3}{2}$ can be attributed to the negative sign of $\mathcal{H}_{twist}$ in Eq. (\ref{twi}), which in turn reflects a new effect of the twist term, and might belong to the SU(2)$_{k=2}$ WZWN universality class. The RG flows in this frustrated three-leg Heisenberg spin tube with $\emph{J}_{3}$ $\gg$ $\emph{J}_{2}$ differ from those of the two-leg zigzag spin ladder.

\acknowledgments

During the course of this work, we have benefitted from very useful discussions with Fabian H. L. Essler, A. A. Nersesyan, De-Shan Yang, Zheng-Chuan Wang, Qing-Rong Zheng, and D. N. Sheng. This work is
supported in part by the NSFC (Grant Nos. 90922033 and 10934008), the MOST of China (Grant
No. 2012CB932901, No. 2013CB933401), and the CAS. We also thank the US NSF Grants DMR-0906816 and DMR-1205734.

\end{document}